\documentclass[preprint,12pt]{elsarticle}

\usepackage{amssymb}
\usepackage{amsmath}
\usepackage{subcaption}
\journal{Acta Materialia}

\begin{document}

\begin{frontmatter}



\title{Thermomechanical conversion in metals: dislocation plasticity model evaluation of the Taylor-Quinney coefficient}


\author[ad1]{Charles K. C. Lieou}
\author[ad2]{Curt A. Bronkhorst\corref{cor1}}
\cortext[cor1]{Corresponding author}
\ead{cbronkhorst@wisc.edu}
\address[ad1]{Theoretical Division, Los Alamos National Laboratory, Los Alamos, NM 87545, USA\fnref{laur}}
\address[ad2]{Department of Engineering Physics, University of Wisconsin-Madison, Madison, WI 53706, USA}
\fntext[laur]{LA-UR release number: LA-UR-20-24985}

\begin{abstract}
Using a partitioned-energy thermodynamic framework which assigns energy to that of atomic configurational stored energy of cold work and kinetic-vibrational, we derive an important constraint on the Taylor-Quinney coefficient, which quantifies the fraction of plastic work that is converted into heat during plastic deformation. Associated with the two energy contributions are two separate temperatures -- the ordinary temperature for the thermal energy and the effective temperature for the configurational energy. We show that the Taylor-Quinney coefficient is a function of the thermodynamically defined effective temperature that measures the atomic configurational disorder in the material. Finite-element analysis of recently published experiments on the aluminum alloy 6016-T4 \citep{neto_2020}, using the thermodynamic dislocation theory (TDT), shows good agreement between theory and experiment for both stress-strain behavior and temporal evolution of the temperature. The simulations include both conductive and convective thermal energy loss during the experiments, and significant thermal gradients exist within the simulation results. Computed values of the differential Taylor-Quinney coefficient are also presented and suggest a value which differs between materials and increases with increasing strain.
\end{abstract}

\begin{keyword}
Constitutive behavior \sep Taylor-Quinney coefficient \sep Thermomechanical conversion \sep Thermodynamics \sep Thermodynamic dislocation theory \sep Dislocation plasticity \sep Aluminum alloy


\end{keyword}

\end{frontmatter}


\section{Introduction}
\label{sec:1}

This manuscript is concerned with a topic which has been explored experimentally for many decades now, namely, the specific partition of deformation energy within the deforming material. This area has been extensively reviewed under the heading of ``The stored energy of cold work'' in two publications \citep{titchener_1958,bever_1973} and covers work through the early 1970's. The authors, in addition to providing a critical review of all the relevant work up to that period of time, also offered significant thought to the likely physical mechanisms responsible for the portion of work stored within the material. They were very clear to point out that the structural evolution which leads to hardening behavior differs with material, loading conditions, and temperature; as such, these factors also impact the energy storage mechanisms. They suggested the potential importance of developed dislocation subgrain structure as a very common storage mechanism. Other possible important variables identified were extent and rate of deformation, grain size, single/polycyrstal, elemental or alloy, purity, and recrystallization. They also openly recognized that measurements related to quantifying stored energy were challenging and therefore different measurement methods would possibly lead to different results. As a result, one may expect a high degree of variability in results with replication advisable within studies. The authors encouraged the readers that ``the ultimate aim of theoretical analysis must be the development of a detailed model of the cold-worked structure resulting from the deformation process. The stored energy, as a measure of the integrated effects of imperfections, will then be predictable from theory''. This statement still very much holds true as we are still challenged to quantitatively understand and represent the integrated structural and thermodynamic evolution process. With this inspirational history of observation this topic remains very rich for discovery with important strategic and economic implications.

The theme of stored energy of cold work has lived on in a more recent body of work examining the evolution of thermal energy through temperature measurement during deformation. A nice review of this work and introduction of new work has been presented recently \citep{rittel_2017} in the context of the Taylor-Quinney coefficient \citep{taylor_1934}. The database covers a span of nine different materials and a number of differing loading conditions \citep{Kapoor_1998,Mason_1994,Macdougall_2000,Rittel_2008,Hodowany_2000,Xia_1990,Rittel_2007,Ghosh_2017}. Their review included a summary of ranges of value for the Taylor-Quinney coefficient using either the integral or differential measure. The range of coefficient values was very large. The lowest magnitude range was 0.1 - 0.3 reported by \citep{Ghosh_2017} for Mg deformed in shear at dynamic rates of loading. The highest magnitude range was 0.75 - 1.0 reported by \citep{Hodowany_2000} for commercially pure Ti loaded dynamically in compression. Within all the measured ranges for each study, the difference between minimum and maximum reported coefficient was on average on the order of 0.4. This is a wide variation and certainly much broader than commonly believed and typically used in engineering practice (e.g., constant 0.9) . The authors themselves \citep{rittel_2017} reported on experiments conducted on seven different materials in a number of initial conditions for the loading modes of compression, tension, and dominant shear. Once again, the results presented suggested values of the Taylor-Quinney coefficient to be varied but certainly well below the traditional value of 0.9. Some additional work to note is that of \citep{rittel_2012} examining the deformation response of single crystal Cu for various loading orientations. The temperature evolution during the dynamic compression loading of 316L SS was reported as a function of strain rate, initial temperature, and strain magnitude in \citep{Lee_2011}. The authors report a comprehensive collection of conditions with full reporting on stress response for each experiment. They report a temperature rise of 140 $^{\circ}$C at a strain of 0.4 deformed at a strain rate of 5000 s$^{-1}$ initially at 25 $^{\circ}$C. Ref.~\citep{Rusinek_2009} experimentally examined TRIP steel under tension using IR thermometry with combined plasticity and phase transformation and demonstrated the influence of the additional energy exerted by the transformation process. They also present a comprehensive analysis of their results for this coupled deformation process which is quite interesting.  Four materials (commercially pure Ti, 303 and 316 SS, Ti-6Al-4V) are studied experimentally by \citep{Knysh_2015} using a combination of DIC and IR thermometry. Austenitic steel was studied by \citep{Oliferuk_1996}, where comparison between fine-grained and coarse-grained material of the amount of stored energy at strains less than a few per cent suggested that the fine-grained material stored more energy than the coarse-grained material. This is consistent with some of the work summarized by \citep{titchener_1958,bever_1973}. This study also examined the contribution of heat loss by conduction and radiation for quasi-static rates of loading and the influence on the measured results so they were also careful to minimize these losses in the design of their experiments. Ref. \citep{Zhang_2018} use IR thermometry to examine the Split-Hopkinson Pressure Bar response of Al 7075-T651 for samples deformed at the rates of 1100 - 4200 s$^{-1}$. Their results suggested that the Taylor-Quinney coefficient is a strong function of strain rate for this material with the coefficient increasing with strain rate. The studies of both \citep{Pottier_2013} and \citep{Fekete_2015} performed integrated experimental and computational studies to give comprehensive descriptions of the experimental results. Finally, very recently \citep{neto_2020} published experimental results of tension of Al 6016-T4 using thermocoupled samples under varying conditions and material ages. We will discuss these results more throughout this manuscript. This collection of work represents an important database for both inspiration and challenge.

As alluded to above by the statement by \citep{titchener_1958,bever_1973}, quantitative explanation of these rich experimental results is an important challenge. Quantifying the stored energy of cold work will not only require accurate description of the mechanics of dislocation motion and interaction \citep{benzerga_2005,longere_2008b}, but must also be done within a framework of thermodynamic consistency \citep{rosakis_2000,anand_2015,zubelewicz_2019} to enable connection with experiment. A great deal of progress has been made over the years to represent the complex mechanics of dislocations within a continuum context \citep{Arsenlis_1999,Arsenlis_2002,Gurtin_2000,Gurtin_2005,Acharya_2004,Anand_2005,Busso_1996,Busso_2000,Aifantis_1992,Fleck_1993,Fleck_1997,Zhu_1995,Gerken_2008,Mayeur_2011,Mayeur_2015}. These important advancements in theory have been greatly facilitated by large-scale dislocation physics simulations to enable the study of details not yet available experimentally \citep[e.g.,][]{KubinBook_2013,BulatovBook_2007,Bulatov_2017,Madec_2002,Madec_2003,Devincre_2006,Devincre_2008,Grilli_2018,dequiedt_2015,hansen_2013}. There has also been significant accomplishment towards understanding the thermodynamics of these dislocation motion and interaction processes  \citep{Berdichevsky_2006,Berdichevsky_2017,Berdichevsky_2018,Berdichevsky_2018a,Berdichevsky_2019,Berdichevsky_2019a,anand_2015,Le_2018b,Le_2019a,Le_2019b,Hochrainer_2016,Levitas_2015a,Arora_2019,Po_2019,Chowdhury_2019,Jiang_2019,nietofuentes_2018,Jafari_2017,Shizawa_2001,RiveraDiazdelCastillo_2012,langer_2010,langer_2015,Roy_2005,Roy_2006,Acharya_2010}  . We have previously proposed a novel thermodynamic description of structural evolution taking place during deformation \citep{lieou_2018,lieou_2019}. This approach is based upon recent work to partition energy and entropy elements to both atomic vibrational and configurational disorder states of matter \citep{langer_2010,langer_2015}. This general theory, based upon principles of the non-equilibrium statistical physics of the collective behavior of driven many-particle systems, then provides a disciplined thermodynamic framework to describe structural evolutionary features such as dislocations, dislocation subcells, grain boundaries, and so on. This is where in this theory the stored energy of cold work resides and is made explicit and distinct from the atomic vibrational component which is responsible for the thermal energy. In our prior work \citep{lieou_2018,lieou_2019} and other recent applications of this theory \citep{langer_2016,langer_2017a,langer_2017b,le_2018a,Le_2018b,Le_2019a,Le_2019b} the atomic configurational disorder component or dislocation, subcell, grain boundary representation was done through isotropic and scalar representations of these states. Here we advance upon the work in \citep{lieou_2018,lieou_2019} by exploring the behavior of the Al alloy 6016-T4 and the experimental database recently published by \citep{neto_2020}. We do so in order to better quantify the stored energy of cold work within the context of a series of very recently published quasi-static experimental results and challenge the ideas within the theory in isotropic form. We have also recently extended the theory to crystallographic systems to represent single crystals \citep{lieou_2020}.

This paper is structured as follows. In Sec.~\ref{sec:2}, we present the thermodynamic framework and its development to derive the appropriate expression for the Taylor-Quinney coefficient. The theory specific to dislocation mechanics is discussed in Sec.~\ref{sec:3}. By employing the theory with implementation into ABAQUS Standard, we present our analysis of the experimental results of \citep{neto_2020} with related discussion.  We then present some concluding thoughts in Sec.~\ref{sec:5}.

\section{Thermodynamic framework: derivation of the Taylor-Quinney coefficient}
\label{sec:2}

In this section, we derive constraints for the Taylor-Quinney coefficient using partitioned-energy thermodynamics \citep{bouchbinder_2009b}. We do not specialize ourselves to dislocation plasticity until the next Section.

The basic premise is that the formation energy of defects in a material $e_D$ far exceeds the thermal energy corresponding to temperature $\theta \equiv k_B T$, and that irreversible atomic rearrangements that correspond to the motion of defects occur on a much longer time scale than the atomic vibration period $\tau_0$, such that one can partition the total energy and entropy per unit volume, $U_{\text{tot}}$ and $S_{\text{tot}}$, into a sum of atomic configurational (C) and kinetic-vibrational (K) contributions. The configurational energy component represents the stored energy of cold work as we discussed above. The same goes for their rates of change:
\begin{equation}
 \dot{U}_{\text{tot}} = \dot{U}_C + \dot{U}_K ; \quad \dot{S}_{\text{tot}} = \dot{S}_C + \dot{S}_K .
\end{equation}
Defects belong to the configurational degrees of freedom, and contribute to energy $U_C$ and entropy $S_C$. (We do not currently consider low-energy defects or other microstructural changes, e.g., phase changes, that pertain to the kinetic-vibrational degrees of freedom; these can be included easily.) The thermal and effective temperatures, denoted respectively by $\theta$ and $\chi$, are defined by
\begin{equation}
 \theta \equiv \dfrac{\partial U_K}{\partial S_K} ; \quad \chi \equiv \dfrac{\partial U_C}{\partial S_C} .
\end{equation}
Then
\begin{equation}\label{eq:Ukdot}
 \dot{U}_K = \theta \dot{S}_K - p \dfrac{\dot{V}^{\text{el}}}{V},
\end{equation}
where $p$ is the hydrostatic pressure, $V$ is the total volume, and $\dot{V}^{\text{el}}$ is the rate of change of the elastic volume. Also,
\begin{equation}\label{eq:Ucdot}
 \dot{U}_C = \chi \dot{S}_C + \sum_{\alpha} \left( \dfrac{\partial U_C}{\partial \rho_{\alpha}} \right)_{S_C,t} \dot{\rho}_{\alpha} + \dfrac{\partial U_C}{\partial t},
\end{equation}
where $\alpha$ indexes the internal state variables $\rho_{\alpha}$ that describe the defects and structural features in the material. These could include, for example, dislocation subcells, the dislocation density $\rho$ and the grain boundary density $\xi$ in a polycrystalline material \citep{lieou_2018,lieou_2019}.

Now, the first law of thermodynamics says that the rate of change of the internal energy per unit volume $\dot{U}_{\text{tot}}$ is equal to the rate of work minus the heat loss to the surroundings and the neighboring material. Thus,
\begin{equation}\label{eq:firstlaw}
 \dot{U}_{\text{tot}} = \dot{U}_C + \dot{U}_K = \sigma_{ij} \dot{\epsilon}_{ij} - \nabla \cdot \mathbf{q} - \nabla \cdot \mathbf{q}_c - A (\theta - \theta_0) .
\end{equation}
Here, $A$ is a conductivity that describes heat exchange with the surroundings at temperature $\theta_0$. $\sigma_{ij}$ and $\dot{\epsilon}_{ij}$ are the Cauchy stress and total strain rate tensors, the latter assumed to be a simple sum of plastic and elastic strain rates: $\dot{\epsilon}_{ij} = \dot{\epsilon}_{ij}^{\text{pl}} + \dot{\epsilon}_{ij}^{\text{el}}$. The heat fluxes assume the form
\begin{equation}
 \mathbf{q} = - K \nabla \theta ; \quad \mathbf{q}_c = -K_c \nabla \chi ,
\end{equation}
where $K$ and $K_c$ are the respective heat conductivities within the material. Substituting Eqs.~\eqref{eq:Ukdot} and \eqref{eq:Ucdot} for $\dot{U}_K$ and $\dot{U}_C$ into Eq.~\eqref{eq:firstlaw}, and using the fact that deformation at constant entropy $S_C$ and defect densities $\rho_{\alpha}$ is by definition elastic, i.e.,
\begin{equation}
 \left( \dfrac{\partial U_C}{\partial t} \right)_{S_C, \rho_{\alpha}} = s_{ij} \dot{e}_{ij}^{\text{el}} ,
\end{equation}
where $s_{ij} = \sigma_{ij} - (1/3) \delta_{ij} \sigma_{kk}$ is the deviatoric stress tensor, and $\dot{e}_{ij}^{\text{el}} = \dot{\epsilon}_{ij}^{\text{el}} - (1/3) \delta_{ij} \dot{\epsilon}_{kk}^{\text{el}}$ is the deviatoric part of the elastic strain rate tensor, we arrive at the result
\begin{equation}\label{eq:firstlaw_2}
 \sigma_{ij} \dot{\epsilon}_{ij}^{\text{pl}} = \chi \dot{S}_C + \sum_{\alpha} \left( \dfrac{\partial U_C}{\partial \rho_{\alpha}} \right)_{S_C, t} \dot{\rho}_{\alpha} + \theta \dot{S}_K - K \nabla^2 \theta - K_c \nabla^2 \chi + A (\theta - \theta_0) . ~~~~~
\end{equation}

Meanwhile, the second law of thermodynamics mandates non-negative entropy production; thus,
\begin{equation}\label{eq:secondlaw}
 \dot{S}_K + \dot{S}_C + \nabla \cdot \dfrac{\mathbf{q}}{\theta} + \nabla \cdot \dfrac{\mathbf{q}_c}{\chi} + \dfrac{A}{\theta} ( \theta - \theta_0 ) \geq 0 .
\end{equation}
Multiplying this by $\chi$, and using Eq.~\eqref{eq:firstlaw_2} for $\chi \dot{S}_C$, we find for the dissipation rate that
\begin{eqnarray}\label{eq:Wc}
 \nonumber \sigma_{ij} \dot{\epsilon}_{ij}^{\text{pl}} - \sum_{\alpha} \left( \dfrac{\partial U_C}{\partial \rho_{\alpha}} \right)_{S_C, t} \dot{\rho}_{\alpha} + (\chi - \theta) \left[ \dot{S}_K - \dfrac{K}{\theta} \nabla^2 \theta + \dfrac{A}{\theta} (\theta - \theta_0) \right] \\ + \dfrac{\chi}{\theta^2} K (\nabla \theta)^2 + \dfrac{K_c}{\chi} ( \nabla \chi )^2 \geq 0 . 
\end{eqnarray}
Assuming plastic incompressibility, $\sigma_{ij} \dot{\epsilon}_{ij}^{\text{pl}} = s_{ij} \dot{\epsilon}_{ij}^{\text{pl}}$, and isotropic plasticity, $\dot{\epsilon}_{ij} \propto s_{ij}$, we arrive at the two constraints
\begin{eqnarray}
 - \left( \dfrac{\partial U_C}{\partial \rho_{\alpha}} \right)_{S_C, t} \dot{\rho}_{\alpha} & \geq & 0, \quad \forall \alpha ; \\
 (\chi - \theta) \left[ \dot{S}_K - \dfrac{K}{\theta} \nabla^2 \theta + \dfrac{A}{\theta} (\theta - \theta_0) \right] & \geq & 0 .
\end{eqnarray}
(The other three inequalities that follow from Eq.~\eqref{eq:Wc} are automatic.) The second of these, in particular, stipulates that
\begin{equation}\label{eq:B}
 \theta \dot{S}_K - K \nabla^2 \theta + A (\theta - \theta_0) = B (\chi - \theta) \approx B \chi ,
\end{equation}
where $B$ is a non-negative thermal transport coefficient, and the approximation follows from the assumption that $\chi \gg \theta$.

Equation ~\eqref{eq:B} provides an important constraint on the Taylor-Quinney coefficient. To see this, consider the non-equilibrium steady-state -- in the long-time limit -- at which the effective temperature has reached some steady-state value $\chi_{\text{ss}}$, and $\dot{\chi} = 0$, $\dot{\rho}_{\alpha} = 0$. Then, substituting Eq.~\eqref{eq:B} into Eq.~\eqref{eq:firstlaw_2}, we find that
\begin{equation}\label{eq:B_req}
 B \rightarrow \dfrac{\sigma_{ij} \dot{\epsilon}_{ij}^{\text{pl}}}{\chi - \theta} , \quad \text{as} \quad t \rightarrow \infty .
\end{equation}
Thermodynamic consistency is guaranteed as long as $B$ satisfies this requirement in the long-time limit. The usual choice, as in, e.g.,~\citep{langer_2015,lieou_2018,lieou_2019}, is
\begin{equation}\label{eq:B1}
 B \equiv B_1 = \dfrac{\sigma_{ij} \dot{\epsilon}_{ij}^{\text{pl}}}{\chi - \theta} .
\end{equation}
However, this is not the only possibility. For example,
\begin{equation}\label{eq:B2}
 B \equiv B_2 = \dfrac{\chi}{\chi_{ss}} \dfrac{\sigma_{ij} \dot{\epsilon}_{ij}^{\text{pl}}}{\chi - \theta} ,
\end{equation}
and more generally,
\begin{equation}\label{eq:B_general}
 B = f(\chi, \theta) \dfrac{\sigma_{ij} \dot{\epsilon}_{ij}^{\text{pl}}}{\chi - \theta} ,
\end{equation}
with $f (\chi_{\text{ss}}, \theta) = 1$, is also thermodynamically consistent.

Then, returning to the equation for the thermal temperature,
\begin{equation}
 \theta \dot{S}_K = \theta \left( \dfrac{\partial S_K}{\partial \theta} \right)_V \dot{\theta} + \theta \left( \dfrac{\partial S_K}{\partial V} \right)_{\theta} \dot{V} \equiv c_V \dot{\theta} + c_V \, \gamma_G \, \theta \, \dot{\epsilon}_{kk} ,
\end{equation}
where $c_V \equiv (\partial S_K / \partial \theta)_V$ is the specific heat capacity per unit volume, and 
\begin{equation}\label{eq:gammaG}
 \gamma_G \equiv \dfrac{V}{c_V} \left( \dfrac{\partial S_K}{\partial V} \right)_{\theta} = - \dfrac{V}{\theta} \left( \dfrac{\partial \theta}{\partial V} \right)_{S_K}
\end{equation}
is the Gr\"{u}neisen parameter, which represents the temperature drop during elastic volume expansion. Then
\begin{eqnarray}\label{eq:thetadot}
 \nonumber c_V \dot{\theta} &=& \theta \dot{S}_K - c_V \, \gamma_G \, \theta \, \dot{\epsilon}_{kk} \\ \nonumber &=& B (\chi - \theta) - c_V \, \gamma_G \, \theta \, \dot{\epsilon}_{kk} + K \nabla^2 \theta - A (\theta - \theta_0) .
\end{eqnarray}
In the case that $B = B_1$ given in Eq.~\eqref{eq:B1}, we can read off the Taylor-Quinney coefficient $\beta$, or the fraction of plastic work dissipated as heat, immediately from Eq.~\eqref{eq:thetadot}:
\begin{equation}\label{eq:beta1}
 \beta \equiv \beta_1 = \dfrac{\chi - \theta}{\chi_{\text{ss}} - \theta} \approx \dfrac{\chi}{\chi_{\text{ss}}} .
\end{equation}
If $B = B_2$, where $B_2$ is given in Eq.~\eqref{eq:B2}, then
\begin{equation}\label{eq:beta2}
\beta \equiv \beta_2 = \dfrac{\chi}{\chi_{\text{ss}}}  \dfrac{\chi - \theta}{\chi_{\text{ss}} - \theta} \approx \left( \dfrac{\chi}{\chi_{\text{ss}}} \right)^2 .
\end{equation}
Both possibilities are thermodynamically consistent, and as we remarked above, there are many possibilities which are second-law compatible; the only requirement is Eq.~\eqref{eq:B_req}, which implies that $\beta$ is a function of the effective temperature, with $\beta \rightarrow 1$ in the long-time limit. Importantly, this implies that the common assumption that $\beta = 0.9$ throughout the duration of deformation cannot be correct, as it is not thermodynamically consistent. Physical evidence found in experimental results will help guide the correct functional form of $\beta$ which is what we do here for a single material in aluminum alloy 6016-T4.

The choice of $\beta$ has implications for the evolution of the effective temperature $\chi$; indeed, plastic work that is not converted into heat is stored in the form of defects and configurational disorder. To write down the evolution equation for $\chi$, use Eqs.~\eqref{eq:firstlaw_2} and \eqref{eq:B} to write
\begin{equation}\label{eq:Scdot}
 \chi \dot{S}_C = \sigma_{ij} \dot{\epsilon}_{ij}^{\text{pl}} - \sum_{\alpha} \left( \dfrac{\partial U_C}{\partial \rho_{\alpha}} \right)_{S_C,t} - B (\chi - \theta) + K_c \nabla^2 \chi .
\end{equation}
Using
\begin{eqnarray}
 \label{eq:U_0} U_C (S_C , \rho_{\alpha}) &=& U_0 (\rho_{\alpha} ) + U_1 (S_1) ; \\
 \label{eq:S_0} S_C (\rho_{\alpha}) &=& S_0 (\rho_{\alpha} ) + S_1 (U_1) ,
\end{eqnarray}
where $U_0$ and $S_0$ denote the total formation energy and bare entropy of the defects, and $U_1$ and $S_1$ denote the residual terms. Next, write
\begin{eqnarray}\label{eq:chidot0}
 \nonumber \chi \dot{S}_C &=& \chi \left( \dfrac{\partial S_C}{\partial \chi} \right)_{\rho_{\alpha}} \dot{\chi} + \chi \sum_{\alpha} \left( \dfrac{\partial S_C}{\partial \rho_{\alpha}} \right)_{\chi}  \dot{\rho}_{\alpha} \\ &=& c_{\text{eff}} \dot{\chi} + \chi \sum_{\alpha} \dfrac{\partial S_0}{\partial \rho_{\alpha}} \dot{\rho}_{\alpha} ,
\end{eqnarray}
where we have defined $c_{\text{eff}} \equiv \chi ( \partial S_C / \partial \chi )_{\rho_{\alpha}}$ to be the effective heat capacity. We also use Eq.~\eqref{eq:U_0} to write
\begin{equation}\label{eq:F_C}
 \left( \dfrac{\partial U_C}{\partial \rho_{\alpha}} \right)_{S_C, t} = \dfrac{\partial U_0}{\partial \rho_{\alpha}} - \chi \dfrac{\partial S_0}{\partial \rho_{\alpha}} .
\end{equation}
Combining Eqs.~\eqref{eq:Scdot}, \eqref{eq:chidot0}, \eqref{eq:F_C} and \eqref{eq:thetadot}, we find
\begin{equation}
 c_{\text{eff}} \dot{\chi} = \sigma_{ij} \dot{\epsilon}_{ij}^{\text{pl}} - \sum_{\alpha} \dfrac{\partial U_0}{\partial \rho_{\alpha}} \dot{\rho}_{\alpha} - B (\chi - \theta) + K_c \nabla^2 \chi .
\end{equation}
As such, if $B = B_1$ given in Eq.~\eqref{eq:B1} above, corresponding to $\beta = \beta_1$ in Eq.~\eqref{eq:beta1}, the evolution equation of the effective temperature reads
\begin{equation}\label{eq:chidot1}
 c_{\text{eff}} \dot{\chi} = \sigma_{ij} \dot{\epsilon}_{ij}^{\text{pl}} \left( 1 - \dfrac{\chi}{\chi_{\text{ss}}} \right) - \sum_{\alpha} \dfrac{\partial U_0}{\partial \rho_{\alpha}} \dot{\rho}_{\alpha} + K_c \nabla^2 \chi .
\end{equation}
This is the usual form of the effective-temperature evolution equation in much of the literature \citep{bouchbinder_2009b,langer_2010,langer_2015,Le_2017a,Le_2017b,lieou_2018,lieou_2019}. In other words, the usual effective-temperature evolution equation is equivalent to the statement that $\beta = \beta_1 = \chi / \chi_{\text{ss}}$. An interesting possibility arises if $B = B_2$ and $\beta = \beta_2 = (\chi / \chi_{\text{ss}})^2$, for which the effective temperature evolution equation reads
\begin{equation}\label{eq:chidot2}
 c_{\text{eff}} \dot{\chi} = \sigma_{ij} \dot{\epsilon}_{ij}^{\text{pl}} \left[ 1 - \left( \dfrac{\chi}{\chi_{\text{ss}}} \right)^2 \right] - \sum_{\alpha} \dfrac{\partial U_0}{\partial \rho_{\alpha}} \dot{\rho}_{\alpha} + K_c \nabla^2 \chi .
\end{equation}
Yet other possibilities may arise (Eq.~\eqref{eq:B_general}). In the rest of this manuscript, we shall restrict ourselves to $\beta = \beta_1 = \chi / \chi_{\text{ss}}$ and $B = B_1$, with Eq.~\eqref{eq:chidot1} as our evolution equation for the effective temperature. This is found to fit the stress-strain behavior of the aluminum alloy 6016-T4 much better than, say, $\beta = \beta_2 = (\chi / \chi_{\text{ss}})^2$. The question of the functional dependency of $\beta$ itself may be material-dependent, and calls for further experimental investigations under different loading conditions.

\section{Thermodynamic dislocation theory}
\label{sec:3}

We now specialize to crystalline materials for which dislocations are the primary defects and plasticity carriers of interest. In this section, we briefly review the thermodynamic dislocation theory (TDT) of Langer et al. We restrict ourselves to isotropic plasticity for simplicity. Interested readers may refer to \citep{langer_2010,langer_2015} for details of the development. From here onwards, we scale the effective temperature $\chi$ by the dislocation formation energy $e_D$ so that it becomes dimensionless; we also use as our only internal density variable (collectively denoted by $\{ \rho_{\alpha} \}$ above) the \textit{dimensionless} dislocation density $\rho$, defined in terms of the usual dislocation line length per unit volume $\rho_D$ by $\rho \equiv a^2 \rho_D$, where $a \approx 10 b$ is an atomic length scale roughly equal to ten times the length of the typical Burgers vector, and is related to the minimum separation between dislocations. In addition, we use the temperature in units of Kelvin, $T$, in place of the temperature in energy units, $\theta = k_B T$, used in Sec.~\ref{sec:2}.

Within TDT, the expression for the plastic strain rate is
\begin{equation}\label{eq:TDT_strainrate}
 \dot{\epsilon}_{ij}^{\text{pl}} = \dfrac{1}{2 \tau_0} \dfrac{s_{ij}}{\bar{s}} \sqrt{\rho} \exp \left[ - \dfrac{T_P}{T} e^{- \bar{s} / (\alpha_T \mu \sqrt{\rho} )} \right] .
\end{equation}
The key idea in Eq.~\eqref{eq:TDT_strainrate} is that dislocation motion occurs via thermally-activated depinning, which is subject to a stress barrier proportional to the shear modulus and inversely proportional to the average separation $\propto 1 / \sqrt{\rho}$ between dislocations; and that the average speed of a dislocation line is given by the average separation between dislocations divided by the time spent at a pinning site. Eq.~\eqref{eq:TDT_strainrate} then follows from the Orowan relation, which says that the plastic strain rate is then proportional to both the dislocation density and their average speed. Here, $\tau_0 \sim 10^{-12}$ s is a time scale corresponding to atomic vibrations. $T_P$, in units of Kelvin, quantifies the energetic barrier to depinning, which is tilted by an equivalent shear stress $\bar{s} \equiv \sqrt{(1/2) s_{ij} s_{ij}}$ in accordance with the inner exponential factor. $\alpha_T$ is a constant of the order of 0.1, and $\mu$ is the shear modulus.

The stress evolution equation is given by
\begin{equation}
 \dot{\sigma}_{ij} = 2 \mu (\dot{\epsilon}_{ij} - \dot{\epsilon}_{ij}^{\text{pl}} ) + \lambda \delta_{ij} \dot{\epsilon}_{kk} ,
\end{equation}
where $\lambda$, the first Lam\'e parameter, is related to the Poisson ratio $\nu_0$ and the shear modulus: $\lambda = 2 \mu \nu_0 / (1 - 2 \nu_0) $. Equivalently,
\begin{equation}
 \dot{s}_{ij} = 2 \mu (\dot{e}_{ij} - \dot{\epsilon}_{ij}^{\text{pl}} ) .
\end{equation}
Strain hardening is described by the evolution equation for the dislocation density:
\begin{equation}\label{eq:rhodot}
 \dot{\rho} = \dfrac{\kappa_1}{\bar{\nu}^2} \dfrac{\sigma_{ij} \dot{\epsilon}^{\text{pl}}_{ij}}{\alpha_T \mu} \left( 1 - \dfrac{\rho}{\rho_{\text{ss}}} \right) ,
\end{equation}
where $\kappa_1$ is a constant, and $\bar{\nu} \equiv \log (T_P / T) - \log [ \log ( \sqrt{\rho} / (2 \tau \dot{\bar{e}}) ) ] $, with $\dot{\bar{e}} \equiv \sqrt{ (1/2) \dot{e}_{ij} \dot{e}_{ij} }$ being the equivalent shear rate. Equation \eqref{eq:rhodot} says that the dislocation density changes at a rate proportional to the plastic work rate $\sigma_{ij} \dot{\epsilon}^{\text{pl}}_{ij}$, due to the storage of cold work; $\rho$ approaches, in the long-time limit, the steady-state dislocation density, given by the second law of thermodynamics (cf. Eq.~\eqref{eq:F_C}): $\rho_{\text{ss}} = e^{-1 / \chi}$. The quantity $\bar{\nu}$ controls the hardening rate post-yield.

With the dislocation core energy per unit length being of the order $\mu b^2 / 4 \pi$, the dislocation formation energy scales as $e_D \sim \mu b^2 a / 4 \pi$, so that the evolution equation for the effective temperature (cf. Eq.~\eqref{eq:chidot1}) becomes
\begin{equation}\label{eq:chidot}
 \dot{\chi} = \dfrac{\kappa_{\chi}}{\alpha_T \mu} \left[ \sigma_{ij} \dot{\epsilon}_{ij}^{\text{pl}} \left( 1 - \dfrac{\chi}{\chi_{\text{ss}}} \right) - \dfrac{\mu}{4 \pi} \left( \dfrac{b}{a} \right)^2 \dot{\rho} \right] ,
\end{equation}
where $\kappa_{\chi} \equiv \alpha_T \mu / c_{\text{eff}}$, and we have neglected the conduction term $\sim K_c \nabla^2 \chi$ within the material. Finally, the temperature evolves according to the equation
\begin{equation}\label{eq:Tdot}
 \dot{T} = \dfrac{1}{c_p \bar{\rho}_M} \left[ \dfrac{\chi}{\chi_{\text{ss}}} \sigma_{ij} \dot{\epsilon}_{ij}^{\text{pl}} - A (T - T_0) + K \nabla^2 T \right] - \gamma_G T \dot{\epsilon}_{kk} ,
\end{equation}
where $c_p$ is the specific heat capacity per unit mass, $\bar{\rho}_M$ is the mass density, and $T_0$ is the ambient temperature.

\section{Analysis of experiment on an aluminum alloy}
\label{sec:5}

\begin{figure}
\begin{center}
\includegraphics[scale=0.3]{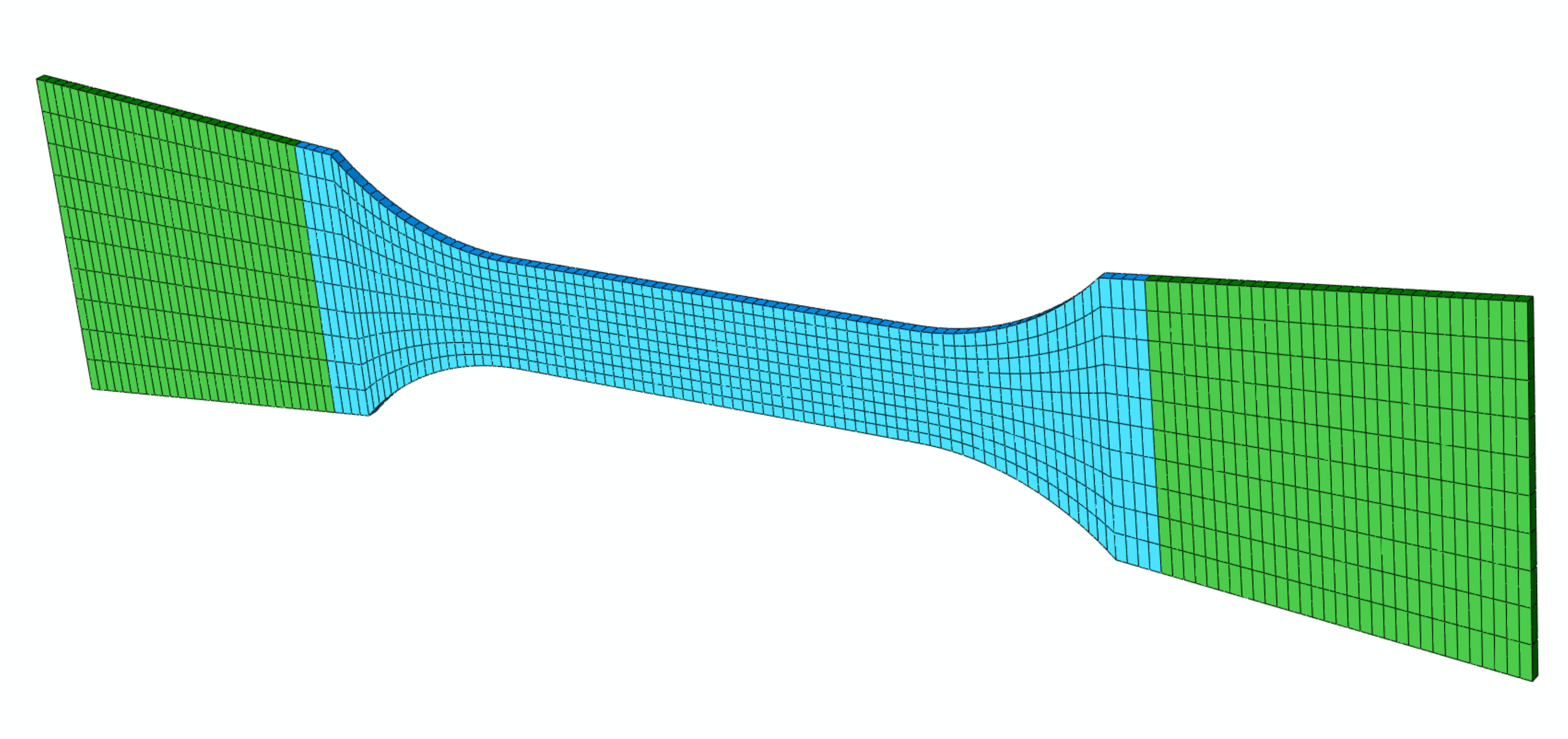}
\caption{\label{fig:Mesh_Sets}The computational ABAQUS mesh used for this work. The green elements contain conductive boundary conditions to represent copper grips used by \citep{neto_2020}. The blue colored elements allow for convective heat transfer to the surrounding air. The left-end nodes are fixed in space while the right end nodes are displaced at rates consistent for each experiment. The linear gage section is 40 mm long and 10 mm wide, and the overall sample length is 144 mm.}
\end{center}
\end{figure}

We perform a finite-element simulation of a set of experiments on the aluminum alloy 6016-T4, reported in detail in \citep{neto_2020}. These experiments measured the evolution of the temperature using thermocouples at four positions on the surface of the aluminum alloy sample deforming under uniaxial tension at several strain rates corresponding to loading velocities $0.01$, $0.1$, and $1$ mm/s at the sample end. These four locations using the designations of \citep{neto_2020} are TC1 - gage section center; TC2 - 6 mm from gage section center; TC3 - 12 mm from gage section center; TC4 - 18 mm from gage section center. The linear gage section is 40 mm long and 10 mm wide. The overall sample length is 144 mm. The computational mesh used for the analysis is given in Figure \ref{fig:Mesh_Sets}. Both thermal conduction into the sample grips and thermal convection to the surrounding air environment are represented in these simulations. The gripped elements with conduction boundary conditions (green) and convection boundary conditions (blue) are shown in Figure \ref{fig:Mesh_Sets}. The nodes on one end of the sample are fixed while the nodes on the opposite end of the model are imposed the experimental displacement rates used in \citep{neto_2020}. Computed temperature evolution is taken from integration points in the model which correspond to the locations of TC1 - TC4.

We integrated the evolution equations in a coupled temperature-displacement analysis in the implicit branch of ABAQUS \citep{abaqus_2014}, using C3D20RT reduced-integration elements and the following procedure. At each time step at time $t$ the strain and temperature increments at time $t + \Delta t$ are automatically updated by the ABAQUS UMAT and HETVAL subroutines, while the state variables $\{ \Lambda_{\alpha} \} \equiv \{ \rho, \chi \}$ are updated as follows in iteration $m$ in the outer loop:
\begin{equation}
 \Lambda_{\alpha}^{(m+1)} (t + \Delta t) = \Lambda_{\alpha} (t) + \Delta t \cdot \dot{\Lambda}_{\alpha} ( \sigma_{ij} (t), \Lambda_{\beta} (t) ) .
\end{equation}
The inner-loop iteration for the stress tensor is a Newton iteration as follows. Let $\Delta \epsilon_{ij}$ be the strain increment accrued over time $\Delta t$, and define
\begin{equation}
 R_{ij} ( \sigma_{kl} (t + \Delta t) ) \equiv \sigma_{ij} (t + \Delta t) - \sigma_{ij} (t) - 2 \mu \left[ \Delta \epsilon_{ij} - \Delta t \, \dot{\epsilon}_{ij}^{\text{pl}} \right] - \lambda \delta_{ij} \Delta \epsilon_{kk} . ~~~~~
\end{equation}
The Newton procedure gives, for the $(n+1)$-st iteration,
\begin{equation}
 \sigma_{ij}^{(n+1)}(t + \Delta t) = \sigma_{ij}^{(n)} (t + \Delta t) - ( J^F_{kl,ij} )^{-1} R_{kl} ( \sigma_{pq}^{(n)} (t + \Delta t) ),
\end{equation}
where 
\begin{equation}
  J_{ij,kl}^F \equiv \dfrac{\partial R_{ij} \left( \sigma_{pq}^{(n)} (t + \Delta t) \right)}{\partial \sigma_{kl}^{(n)} (t + \Delta t)} = \delta_{ij,kl} + 2 \mu \Delta t \dfrac{\partial \dot{\epsilon}^{\text{pl}}_{ij} \left( \sigma_{pq}^{(n)} (t + \Delta t) \right)}{\partial \sigma_{kl}^{(n)} (t + \Delta t)} , 
\end{equation}
with $\delta_{ij,kl}$ being the identity tensor: $\delta_{ij,kl} = 1$ if $(ij) = (kl)$ and 0 otherwise. Upon convergence of the stress update, we return to the outer loop and update the state variables until convergence. The UMAT subroutine makes use of an adaptive time stepping scheme that is allowed to proceed if the temperature change over $\Delta t$ falls within a tolerance of $0.01$ K.

\begin{table}
\scriptsize
\begin{center}
\caption{\label{tab:composition}Composition (by percentage of mass) of the aluminum alloy 6016-T4 used by \citep{neto_2020}}
\begin{tabular}{llllll}
\hline
Si & Mg & Cu & Fe & Mn & Al \\
\hline
0.91 & 0.41 & 0.10 & 0.255 & 0.17 & bal. \\
\hline
\end{tabular}
\end{center}
\end{table}

The composition of the aluminum alloy 6016-T4 is detailed in Table \ref{tab:composition}. It is similar enough to commercial-purity aluminum whose TDT parameters can be found in the literature \citep{Le_2017a,Le_2017b}. We expect that TDT material parameters for the aluminum alloy 6016-T4 should be fairly similar to those for commercially-pure aluminum. To this end, we first use the steady-state flow stress measurements at various temperatures, reported in a separate study of the aluminum alloy \citep{bariani_2014}, to compute the depinning barrier $T_P$ using the procedure described in \citep{langer_2010}, and find that $T_P \approx 24000$ K as in commercially-pure aluminum. The other TDT material parameters are adjusted to fit the stress-strain measurements; these and other known material parameters are tabulated in Table \ref{tab:parameters}. In particular, we assume that the shear modulus $\mu$ is temperature-dependent:
\begin{equation}
 \mu(T) = \mu_1 - \dfrac{D_1}{e^{T_1 / T} - 1} .
\end{equation}
The initial temperature is 295 K; initial dislocation densities and effective temperatures are tabulated in Table \ref{tab:initial}. There are uncertainties with the convective heat transfer coefficients between the sample surface and the surrounding air, and between the sample surface and the grip interface. We assume as in much of \citep{neto_2020} that $A_{\text{air}} = 3$ W m$^{-2}$ K$^{-1}$ and $A_{\text{grip}} = 1750 $ W m$^{-2}$ K$^{-1}$.

\begin{table}
\scriptsize
\begin{center}
\caption{\label{tab:parameters}List of aluminum alloy 6016-T4 material parameters}
\begin{tabular}{lll}
\hline
Parameter & Definition or meaning & Value \\
\hline\hline
$\bar{\rho}_M$ & Mass density & 2800 kg m$^{-3}$ \\
$c_p$ & Specific heat capacity & 850 J kg$^{-1}$ K$^{-1}$ \\
$K$ & Thermal conductivity & 184 W m$^{-1}$ K$^{-1}$ \\
$\mu_1$ & Shear modulus parameter & 28.8 GPa \\
$D_1$ & Shear modulus parameter & 3.44 GPa \\
$T_1$ & Shear modulus parameter & 215 K \\
$\nu_0$ & Poisson ratio & 0.33 \\
$\gamma_G$ & Gr\"{u}neisen parameter & 2.02 \\
$\alpha_T$ & Slip resistance parameter & 0.065 \\
$b$ & Burgers vector & 0.102 nm \\
$a$ & Atomic length scale & 1.02 nm \\
$\tau_0$ & Atomic time scale & 1 ps \\
$T_P$ & Depinning barrier & $24 000$ K \\
$\chi_{\text{ss}}$ & Steady-state effective temperature (in units of $e_D$) & 0.25 \\
$\kappa_1$ & Hardening parameter & 1.15 \\
$\kappa_{\chi}$ & Effective temperature increase rate & 3.5 \\
\hline
\end{tabular}
\end{center}
\end{table}

\begin{table}
\scriptsize
\begin{center}
\caption{\label{tab:initial}List of initial conditions}
\begin{tabular}{llll}
\hline
Sample aging time & Sample edge velocity & $\rho ( t = 0)$ & $\chi (t = 0)$\\
\hline\hline
1 month & 0.01 mm/s & $1.5 \times 10^{-3}$ & 0.215 \\
1 month & 0.1 mm/s & $1.35 \times 10^{-3}$ & 0.207 \\
1 month & 1 mm/s & $1.15 \times 10^{-3}$ & 0.20 \\
18 months & 0.01 mm/s & $2.3 \times 10^{-3}$ & 0.227 \\
18 months & 0.1 mm/s & $2.05 \times 10^{-3}$ & 0.218 \\
18 months & 1 mm/s & $1.85 \times 10^{-3}$ & 0.212 \\
\hline
\end{tabular}
\end{center}
\end{table}

\begin{figure}
\begin{center}
\includegraphics[scale=0.6]{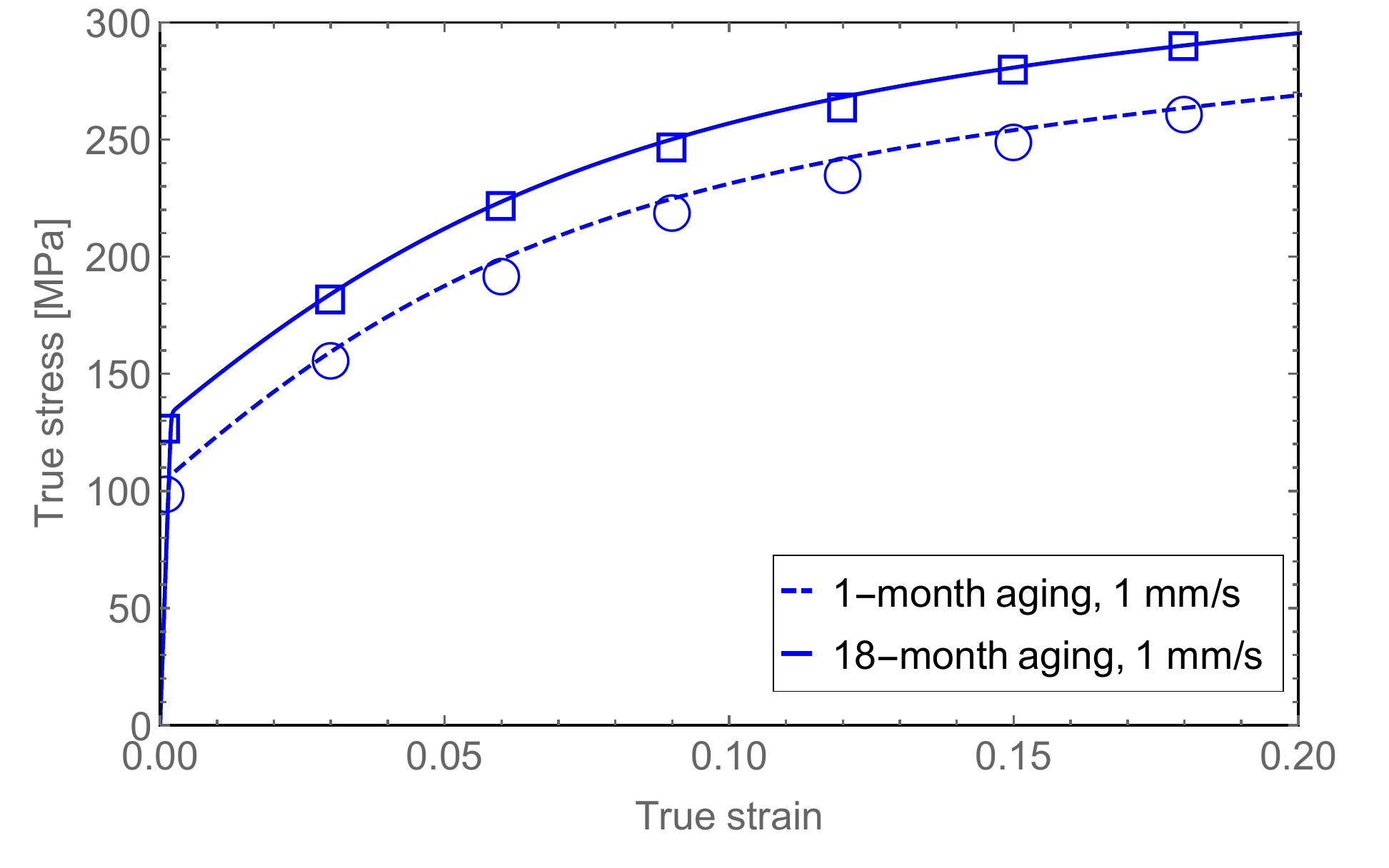}
\caption{\label{fig:stress_all}Stress-strain variation of the aluminum alloy 6016-T4, under uniaxial tension, with edge velocity 1 mm/s corresponding to a nominal strain rate of $2 \times 10^{-2}$ s$^{-1}$. The solid curve shows our theoretical calculations for the alloy aged 18 months, and dashed curve is for the sample aged 1 month. The open squares and circles denote the data taken from \citep{neto_2020}. The stress-strain curves reported in \citep{neto_2020} for the lower strain rates are almost identical, and are omitted here.}
\end{center}
\end{figure}

Figure \ref{fig:stress_all} compares the results of our finite-element calculations for the stress-strain behavior of the aluminum alloy 6016-T4 sample subject to uniaxial tension. We show the results for two samples aged 1 and 18 months, respectively, and only for the highest strain rate reported in \citep{neto_2020} for clarity, as the results for the lower strain rates are almost identical. The stress-strain behavior provides the primary information for us to adjust parameters and initial conditions -- specifically the parameters $\alpha_T$, $\kappa_1$ and $\kappa_{\chi}$ as well as the initial condition $\chi(t = 0)$ which control the curvatures of the stress-strain curves after the onset of strain hardening, and $\rho(t = 0)$ which determines the stress at the onset of strain hardening.

\begin{figure}
\begin{center}
\includegraphics[scale=0.6]{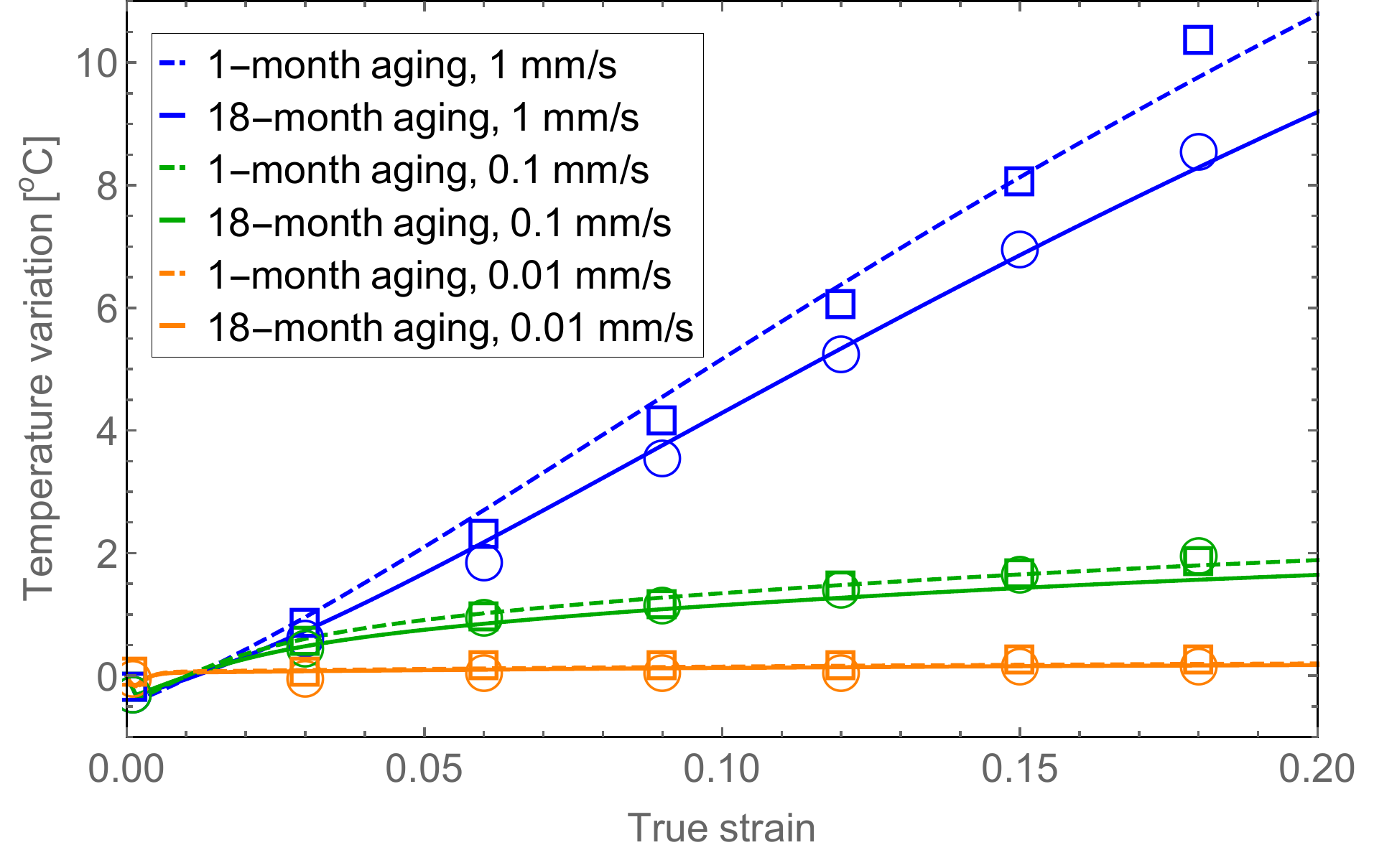}
\caption{\label{fig:temp_all}Temperature variation of the aluminum alloy 6016-T4, under uniaxial tension, with edge velocities 1 mm/s (blue), 0.1 mm/s (green), and 0.01 mm/s (orange), corresponding to a nominal strain rate of $2 \times 10^{-2}$, $2 \times 10^{-3}$, and $2 \times 10^{-4}$ s$^{-1}$. The solid curve shows our theoretical calculations for the alloy aged 18 months, and dashed curve is for the sample aged 1 month. The open squares and circles denote the data taken from \citep{neto_2020}.}
\end{center}
\end{figure}

With good agreement between experimental and theoretical stress-strain curves, we compute the evolution of the temperature at the center of the deforming sample, referred to as position TC1. Figure \ref{fig:temp_all} shows the comparison between the experimental temperature measurements and the results of finite-element calculations. With the same parameters that enabled us to reproduce the stress-strain behavior, we once again demonstrate good agreement between experiment and theory. TDT with a Taylor-Quinney coefficient $\beta_1 = \chi / \chi_{\text{ss}}$ that varies linearly with the effective temperature captures the almost linearly increasing temperature at the center location of the sample, with the discrepancy not exceeding 0.5 K for all strain rates and at all times. The discrepancy may be due to uncertainties in heat convection between the sample and the surrounding air, and heat conduction between the sample and the grip material at its two ends; we do not make further efforts to fit those ingredients. An important observation, however, is that the temperature rise is steepest for the fastest loading rate which, for the same accumulated strain, corresponds to less time available for heat conduction and convection. In addition, by including the elastic thermal expansion term through the Gr\"{u}neisen parameter $\gamma_G$ (Eqs.~\eqref{eq:gammaG} and \eqref{eq:Tdot}) we have been able to accurately describe the small temperature drop of up to about 0.25 K at small strains during elastic deformation, before its subsequent increase due to plasticity-induced heating. While not shown here, we have been unable to fit the temperature increase as closely, or the stress-strain behavior at large strains, if we used $\beta_2 = ( \chi / \chi_{\text{ss}} )^2$ for the Taylor-Quinney coefficient. This suggests that $\beta = \beta_1$ is a plausibility, which calls for further confirmation from other experiments and must be continued on other databases as discussed in the introduction.

\begin{figure}
\begin{center}
\begin{subfigure}[b]{.75\textwidth}
\includegraphics[width=\textwidth]{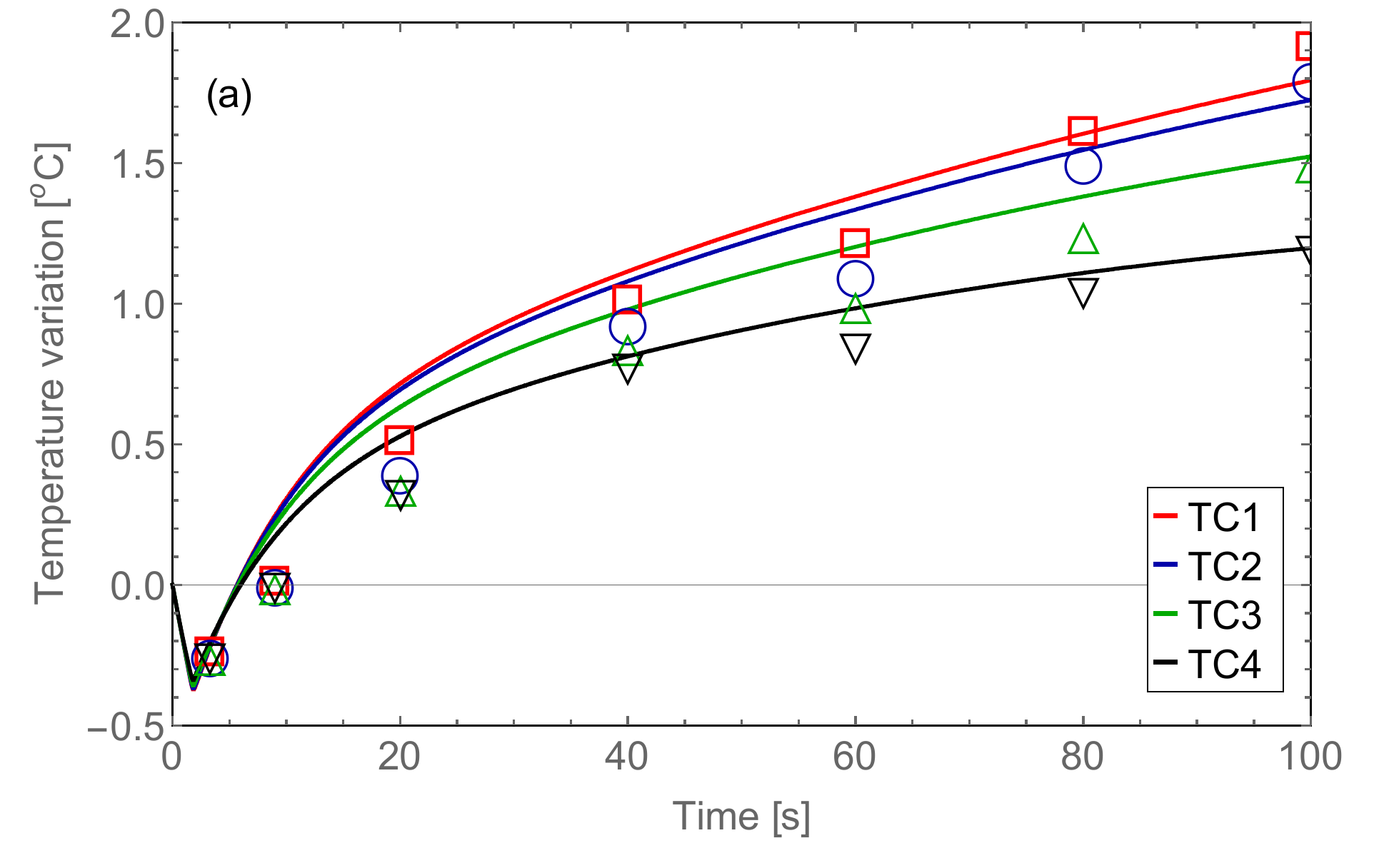}
\end{subfigure}
\begin{subfigure}[b]{.75\textwidth}
\includegraphics[width=\textwidth]{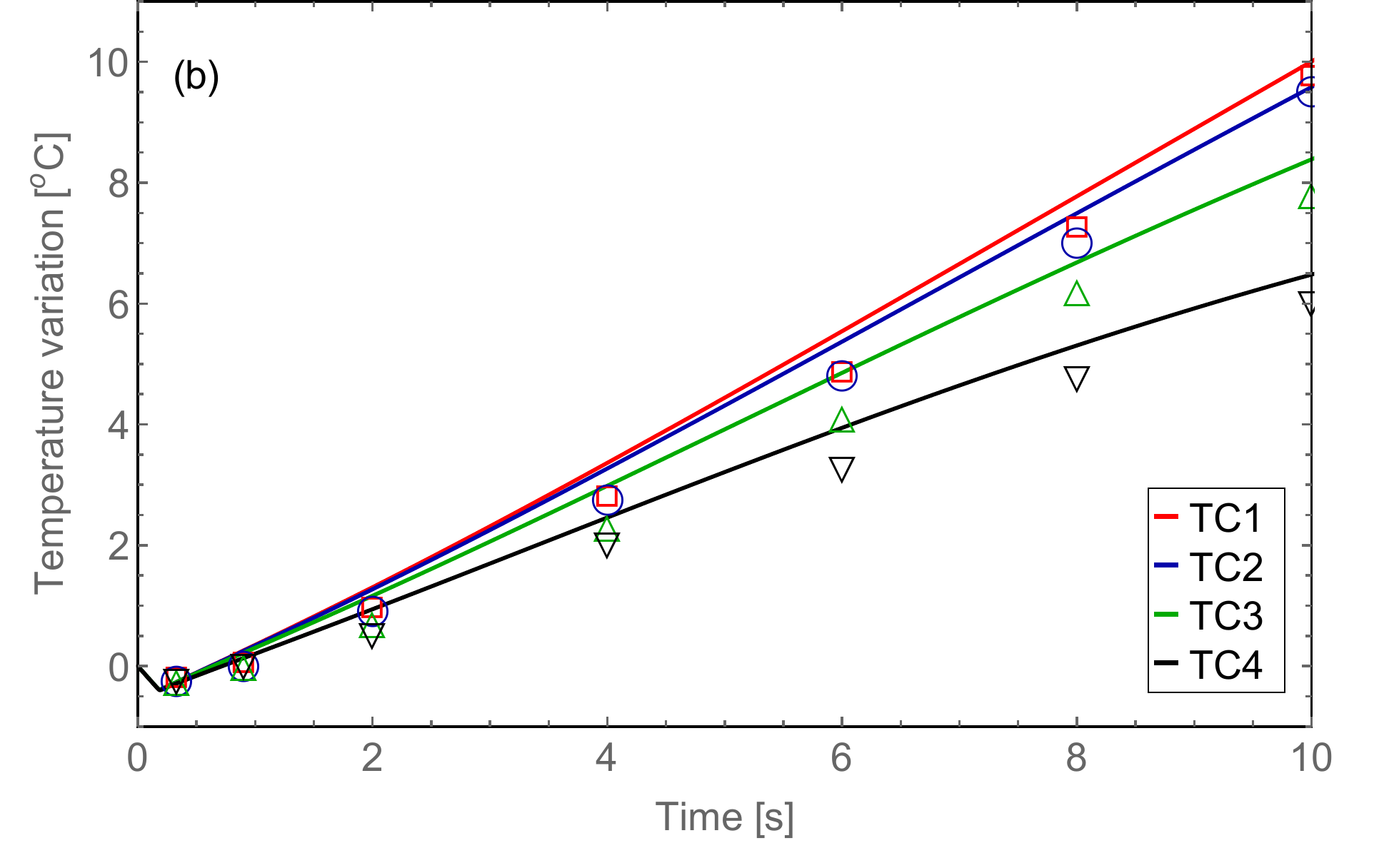}
\end{subfigure}
\caption{\label{fig:temptc}Temperature evolution as a function of time, at several prescribed positions TC1, TC2, TC3, and TC4, for the 6016-T4 aluminum alloy sample aged 18 months, at edge velocities (a) 0.1 mm/s, and (b) 1 mm/s.}
\end{center}
\end{figure}

\begin{figure}
\begin{center}
\includegraphics[scale=0.3]{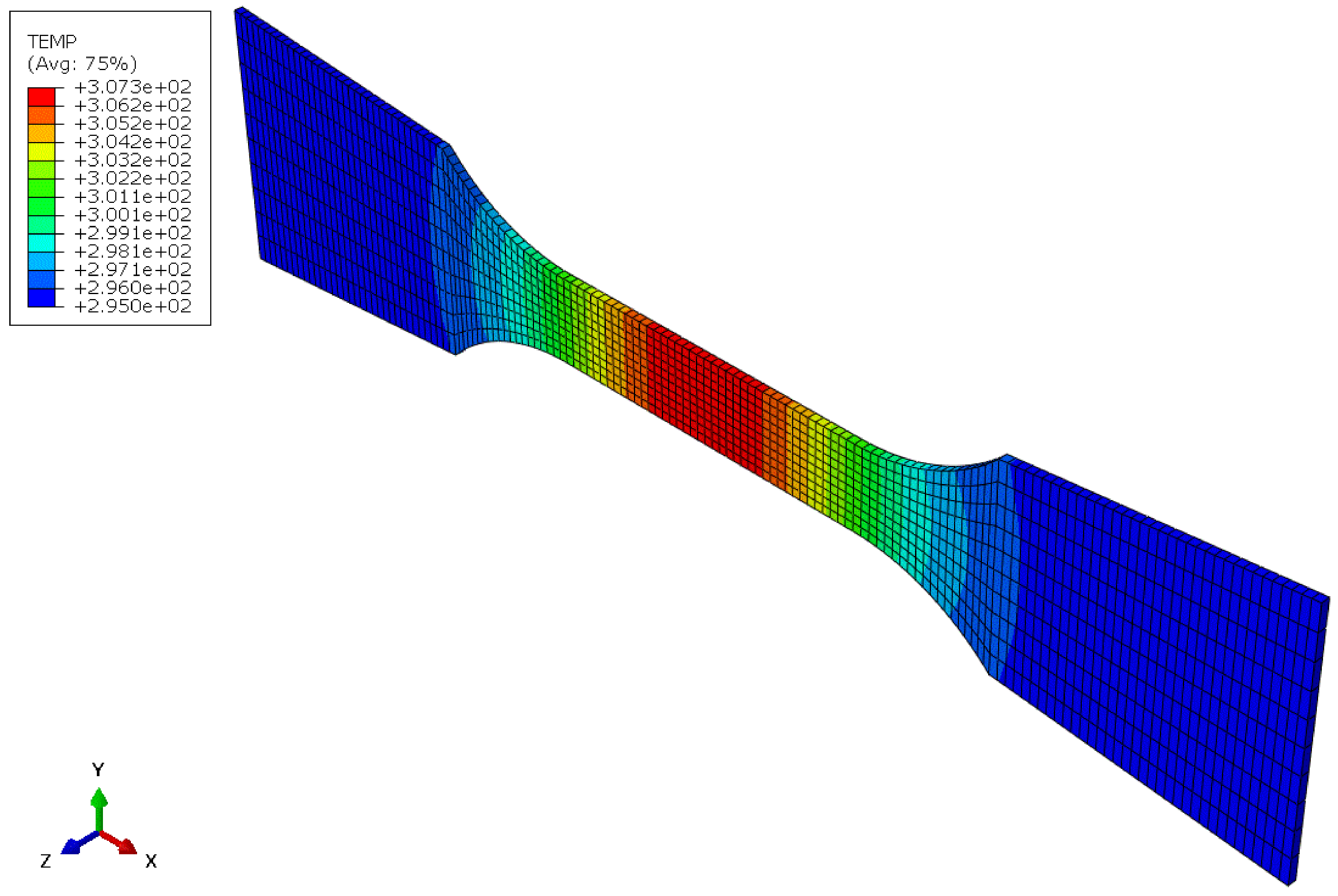}
\caption{\label{fig:mesh_temp}Temperature variation of the tension sample simulation for the case of the 18 month aged material and an end velocity of 1 mm/s. This contour is taken at a simulation time of 10 s.}
\end{center}
\end{figure}

The Neto \textit{et al.}~experiments also reported, in their Fig.~6, measurements of the temperature evolution at several prescribed positions along the long axis of the sample. We compare their measurements with our theoretical results in Fig.~\ref{fig:temptc}. Once again we are able to reproduce the observed temperature increase, with a discrepancy of less than 0.5 K in each case. We attribute the small discrepancy to uncertainties in the heat conduction and convection behavior. Using the computed results for the 18 month aged materials deformed at a displacement velocity of 1 mm/s we show the temperature profile taken at a time of 10 s after the initiation of deformation in Fig.~\ref{fig:mesh_temp}. This result clearly demonstrates the significant influence of conductive heat loss found for these experimental boundary conditions.

\begin{figure}
\begin{center}
\includegraphics[scale=0.6]{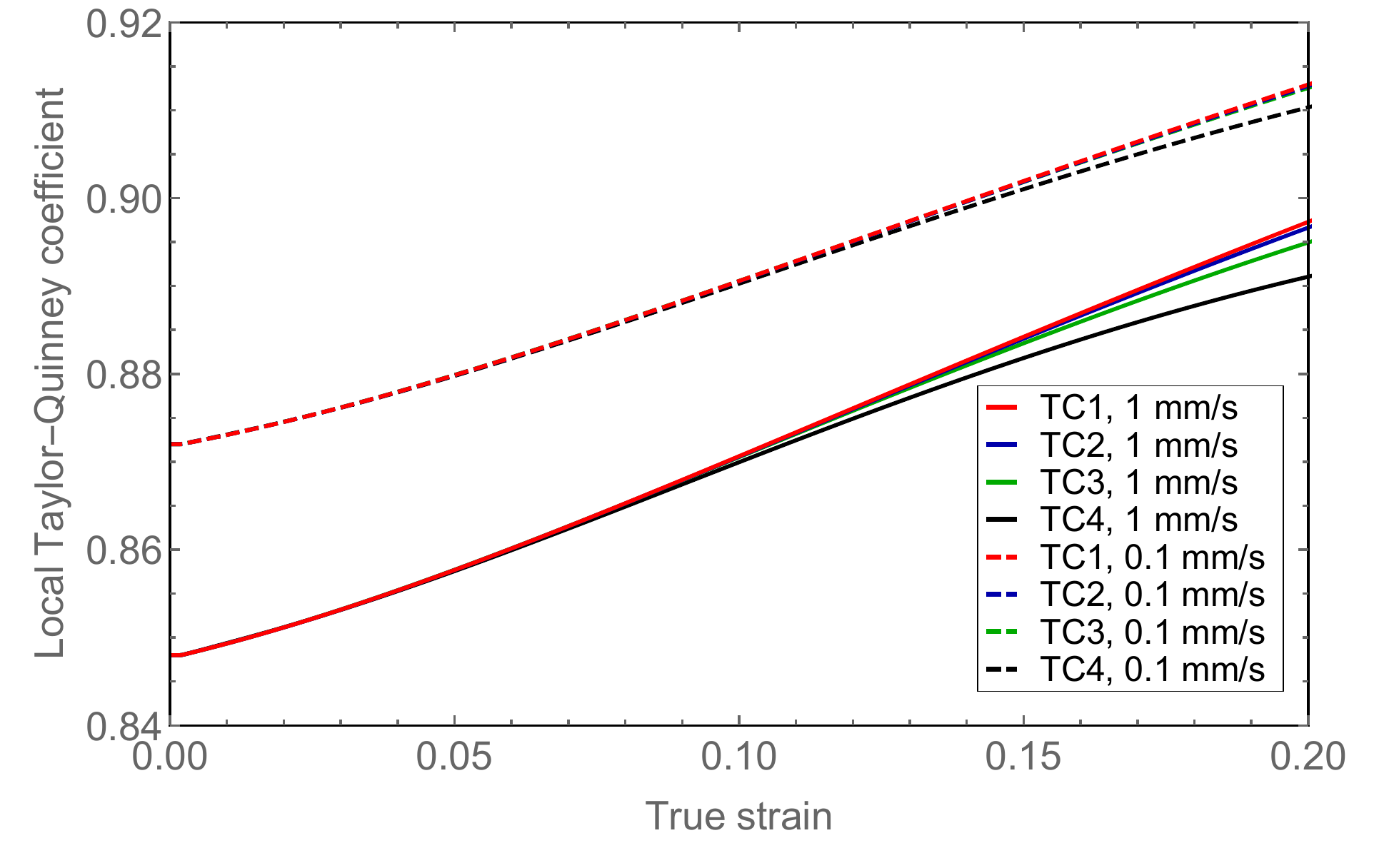}
\caption{\label{fig:betatc}Model results for the variation of the local Taylor-Quinney coefficient $\beta$ as a function of strain, at the prescribed positions TC1, TC2, TC3, and TC4, for the 6016-T4 aluminum alloy sample aged 18 months, at edge velocities 0.1 mm/s and 1 mm/s.}
\end{center}
\end{figure}

Finally, we show in Fig.~\ref{fig:betatc} the model results for the variation of the local Taylor-Quinney coefficient $\beta$, given by Eq.~\eqref{eq:beta1}, as a function of strain, at the same positions as shown in Fig.~\ref{fig:temptc} for the temperature variation. We show only the results for the sample aged 18 months and driven at edge velocities 0.1 mm/s and 1 mm/s; results for other loading rates and the 1 month aged sample are similar. The local Taylor-Quinney coefficient slowly increases with the total strain, and is not a constant; despite the marked differences in the temperature variation shown in Figs.~\ref{fig:temptc} at the different positions, the difference in $\beta$ at these positions is only minor. This indicates that thermal conduction within the sample, and heat convection with the surface and the grip, indeed have a significant effect on the thermal behavior of the deforming material. In other words, deformation is not adiabatic in the present set-up. The fact that we have been able to reproduce both the mechanical and thermal behaviors of the deforming material, despite the complexity of the set-up, suggests that our present model is a reasonably accurate representation of thermomechanical conversion in the 6016-T4 aluminum alloy.

\section{Concluding remarks}
\label{sec:6}

In this paper we have derived constraints on the Taylor-Quinney coefficient $\beta$, which quantifies the fraction of plastic work dissipated as heat, using a thermodynamic framework which partitions deformation energy into elastic, configurational (stored energy of cold work), and kinetic-vibrational. The common Taylor-Quinney coefficient represents coupling between the configurational and vibrational energy states. We have shown that the Taylor-Quinney coefficient is a function of the effective temperature which increases towards unity in the long-time limit. The conventional assumption that $\beta = 0.9$ uniformly over the course of deformation is not thermodynamically consistent. This is also not supported by current experimental evidence and the results of this study using the theory in its current form. We show in addition that $\beta = \beta_1 \equiv \chi / \chi_{\text{ss}}$, where $\chi_{\text{ss}}$ is the effective temperature at the steady state, provides a good fit to experimental measurements in the aluminum alloy 6016-T4 under a range of loading rates and material preparation. Incidentally, this implies that our findings about the Taylor-Quinney coefficient may provide a means to indirectly measure the stored energy of cold work or the effective temperature, a key quantity that quantifies the configurational disorder in the material, and represents the evolution of dislocation structure during deformation. As we alluded to early in the manuscript, development and evolution of dislocation subcells is anticipated to be an important energy storage mechanism and should be explicitly represented. This remains an important area of theoretical development, especially for single-crystal models with important implications for enabling our modeling of hardening and thermodynamic state evolution in metallic materials. In addition, experimental datasets, which include measurement of temperature with the traditional stress and strain curves, offer an important critical piece of information which can enable us to gain a deeper insight into structural evolution processes that can be probed with theories as presented here. Structural-only models ignore an important piece of diagnostic information in temperature which, as we have seen, is significant for even quasi-static rates of loading. 

Important questions remain about the functional dependency of $\beta$ on measurable physical quantities. We cannot definitively conclude that $\beta = \beta_1$; there are other thermodynamically consistent possibilities, whose verification call for further experimentation, as well as atomistic simulations that may be able to provide a direct probe of the stored energy of cold work (effective temperature in the present context). This can be done atomistically via calculations of atomic potential energies and use of fluctuation-dissipation relations, as has been done for amorphous materials~\citep{hinkle_2017,ono_2002} in supplement to experimental information at least qualitatively if not quantitatively. The existing database of experiments covering many different materials is beginning to be built with local temperature evolution with deformation. There are many questions remaining as to the role of dislocation processes and the structural and chemical features of a material which were discussed in the introduction. The creation of new defects and the interactions of evolving defect populations with other defects and chemical and impurity elements within the material defines the problem of material hardening and therefore is an integral part of defining proper theoretical frameworks for the stored energy of cold work. We pose a simple model representing this process, but much more work is required to explicitly represent dominant mechanisms for materials. Potential energy storage mechanisms include those related to dislocation patterning, grain boundary interactions, point defects, alloying elements, dislocation creation, and the atomic lattice distortion-based stored energy related to each. This then necessarily advises us to begin thinking about different models for each material which physically represents the relevant physics given that each material will respond differently to varying stress, loading rate, and temperature conditions. The focus of the discussion here has been strictly with regards to dislocation slip-based processes; however, deformation twinning is also an important plastic deformation mechanism. It is still unclear if, in terms of the stored energy of cold work and thermal conversion, that twinning can be treated the same way as dislocation slip \citep{Kingstedt_2019}.

\section*{Acknowledgements}

Both authors were partially supported by the DOE/DOD Joint Munitions Program and and LANL LDRD Program Project 20170033DR. CAB acknowledges support from the University of Wisconsin Alumni Research Foundation.



\bibliographystyle{elsarticle-num} 
\bibliography{actamat_tq_al_03}


%\begin{thebibliography}{00}


%\bibitem{}

%\end{thebibliography}
\end{document}